\documentclass[onecolumn,prb,notitlepage,floats,numerical,citeautoscript,superscriptaddress,amsmath,amssymb]{revtex4-2}

\usepackage[version=3]{mhchem} 
\usepackage{bm}
\usepackage[utf8]{inputenc}
\usepackage[T1]{fontenc}
\usepackage{graphicx}
\usepackage{upgreek}
\usepackage{color}
\usepackage{ulem}
\usepackage{hyperref} 
\hypersetup{colorlinks,citecolor=blue, filecolor=blue ,linkcolor=blue , urlcolor=blue, pdftex}
\usepackage{sidecap}
\usepackage{amssymb}
\usepackage{siunitx}
\usepackage[caption=false]{subfig}
\usepackage{xcolor}

\def \FUW{Institute of Experimental Physics, Faculty of Physics, University of Warsaw, 02-093 Warsaw, Poland}
\def \SingaporeEng{Department of Materials Science and Engineering, National University of Singapore, 117575, Singapore} 
\def \SingaporeInt{Institute for Functional Intelligent Materials, National University of Singapore, 117544, Singapore} 
\def \Watanabe{Research Center for Electronic and Optical Materials, National Institute for Materials Science, 1-1 Namiki, Tsukuba 305-0044, Japan}
\def \Taniguchi{Research Center for Materials Nanoarchitectonics, National Institute for Materials Science,  1-1 Namiki, Tsukuba 305-0044, Japan}
\def \FUWtheo{Institute of Theoretical Physics, Faculty of Physics, University of Warsaw, 02-093 Warsaw, Poland}
\def \PWr{Department of Semiconductor Materials Engineering, Faculty of Fundamental Problems of Technology, Wroc\l{}aw University of Science and Technology, 50-370 Wroc\l{}aw, Poland}

\usepackage[left=5.4cm, right=1.4cm]{geometry}

\begin{document}

\title{Resonant Raman Scattering of Few Layers CrBr$_3$}

\author{\L{}ucja Kipczak}
\email{lucja.kipczak@fuw.edu.pl}
\affiliation{\FUW}
\author{Arka Karmakar}
\affiliation{\FUW}
\author{Magdalena Grzeszczyk}
\affiliation{\SingaporeInt}
\author{R\'o\.za Janiszewska}
\affiliation{\PWr}
\author{Tomasz Wo\'zniak}
\affiliation{\PWr}
\affiliation{\FUWtheo}
\author{Zhaolong~Chen}
\affiliation{\SingaporeInt}
\affiliation{\SingaporeEng}
\author{Jan Pawłowski}
\affiliation{\FUW}
\author{Kenji~Watanabe}
\affiliation{\Watanabe}
\author{Takashi Taniguchi}
\affiliation{\Taniguchi}
\author{Adam Babi\'nski}
\affiliation{\FUW}
\author{Maciej Koperski}
\affiliation{\SingaporeInt}
\affiliation{\SingaporeEng}
\author{Maciej R. Molas}
\email{maciej.molas@fuw.edu.pl}
\affiliation{\FUW}

\begin{abstract}
We investigate the vibrational and magnetic properties of thin layers of chromium tribromide (CrBr$_3$) with a thickness ranging from three to twenty layers (3~L to 20~L) revealed by the Raman scattering (RS) technique. 
Systematic dependence of the RS process efficiency on the energy of the laser excitation is explored for four different excitation energies: 1.96 eV, 2.21 eV, 2.41 eV, and 3.06 eV. 
Our characterization demonstrates that for 12 L CrBr$_3$, 3.06~eV excitation could be considered resonant with interband electronic transitions due to the enhanced intensity of the Raman-active scattering resonances and the qualitative change in the Raman spectra.
Polarization-resolved RS measurements for 12 L CrBr$_3$ and first-principles calculations allow us to identify five observable phonon modes characterized by distinct symmetries, classified as the A$_\textrm{g}$ and E$_\textrm{g}$ modes.
The evolution of phonon modes with temperature for a 20 L CrBr$_3$ encapsulated in hexagonal boron nitride flakes demonstrates alterations of phonon energies and/or linewidths of resonances indicative of a transition between the paramagnetic and ferromagnetic state at Curie temperature ($T_\textrm{C} \approx 50$ K).
The exploration of the effects of thickness on the phonon energies demonstrated small variations pronounces exclusively for the thinnest layers in the vicinity of 3 - 5 L. 
This observation is attributed to strong localization in the real space of interband electronic excitations, limiting the effects of confinement for resonantly excited Raman modes to atomically thin layers.
\end{abstract}

\maketitle

\section*{Introduction \label{sec:Intro}}
The recent discovery of magnetism in two-dimensional van der Waals (vdW) materials at the limit of atomically thin monolayers opens capabilities to study the fundamental aspects of magnetism at varied dimensionalities. 
From the practical perspective, the realization of magnetism down to monolayers pushes the limits of device minaturization, which becomes increasingly relevant for the development of spintronics~\cite{Jiang2018, Huang2018}, valleytronics, and nanoelectronics~\cite{Gilbertini2019, Wang2022}. 
The family of layered magnetic materials (LMMs) grows rapidly motivated by the search for materials that host stable magnetic orders under a variety of physical conditions, including temperature~\cite{Jin2018, Vaclavkova2020, Yin2021}, magnetic or electric field~\cite{Vaclavkova2020, PhysRevX.10.011075, Jana2023}, and pressure~\cite{Pawbake2022}. 
These intense research efforts lead to theoretical predictions of hundreds of LMM systems, of which dozens have been synthesized and characterized. 
The largest groups of LMMs include di- and trihalides ($e.g.$, CrBr$_3$ and CrI$_3$)~\cite{McGuire2017Crystals, Mak2019}, transition metal dichalcogenides ($e.g.$, 1T-VS$_2$)~\cite{Zhang2013, Wu2023}, tri- and tetrachalcogenides ($e.g.$ FePS$_3$ and CrPS$_4$)~\cite{Takano2004, Lancon2016, Zhang2021, Lee2017, Peng2020, Rybak2023}, and metal-chalcogene-halides ($e.g.$, CrSBr)~\cite{Goser1990, Wang2023}. 

Herein, we investigate the vibrational properties of CrBr$_3$, which belongs to the family of chromium trihalides (CrX$_3$, X = I, Br, Cl). 
The three materials exhibit an intralayer ferromagnetic coupling within a monolayer, however, they differ in the easy-axis magnetization exhibiting in-plane (CrCl$_3$) or out-of-plane spin orientations (CrBr$_3$ and CrI$_3$)~\cite{Gilbertini2019, Soriano2020}. 
The interlayer coupling is more complex, displaying thickness- and/or gate-dependent ferromagnetic or antiferromagnetic characteristics~\cite{Wang2011, Gilbertini2019,McGuire2017Crystals, McGuire2014, Li2020}. 
For CrBr$_3$, the interlayer coupling is mostly ferromagnetic, however, contributions from antiferromagnetic coupling have been observed in bulk crystals via inspection of magneto-resistance in vertical tunnelling junctions~\cite{Yao2023}.

From the perspective of studying lattice dynamics, the Raman scattering (RS) technique has been established as a pivotal tool to uncover the physics of the vibrational and electronic properties of layered vdW materials, as well as a method to identify the exact number of layers~\cite{Lee2010, Grzeszczyk2016, Kipczak2020, Bhatnagar2022}. 
Particularly, the RS characterization of CrBr$_3$ revealed a chiral character of the lattice excitations in a rather complex elementary cell~\cite{Yin2021}. 
These phonon modes are coupled to the magnetic order, predominantly driven by modifications of the exchange coupling between neighboring Cr${^{3+}}$ ions due to vibrational motion of the lattice~\cite{Wu2022}. 

In this work, we further explore the interplay between vibrational, electronic, and magnetic characteristics of thin CrBr$_3$ layers through systematic polarization-resolved RS characterization as a function of the energy of laser excitation, temperature, and the thickness of the magnetic layers. We concluded that for 12 L CrBr$_3$ 3.06~eV excitation energy creates resonant conditions characterized by enhanced RS efficiency at the cryogenic temperature $T$~=~5~K. 
Five Raman peaks were identified in the polarization-sensitive resonant RS spectra, characterized by the A$_\textrm{g}$(out-of-plane) and E$_\textrm{g}$(in-plane) symmetries, as confirmed by first-principles calculations.
Curie temperature ($T_\textrm{C}$) for CrBr$_3$ is determined to be \(\sim\)50~K from the measured sudden energy redshift of the Raman resonances triggered by the increase of temperature above the critical point, where the ferromagnetic order collapses in favor of the paramagnetic response. 
Moreover, the effect of thicknesses on the phonon energies reveals a notable change in phonon energy only for the thinnest layers (3 - 5 L CrBr$_3$), indicative of the strong spatial localization of excitons~\cite{CrBr3_spin_pumping} which are expected to be responsible for the resonant characteristics of the enhanced vibrational response.

\section*{Results and discussion \label{sec:Experimnet}}

\begin{figure}[h]
		\subfloat{}%
		\centering
		\includegraphics[width=0.8\linewidth]{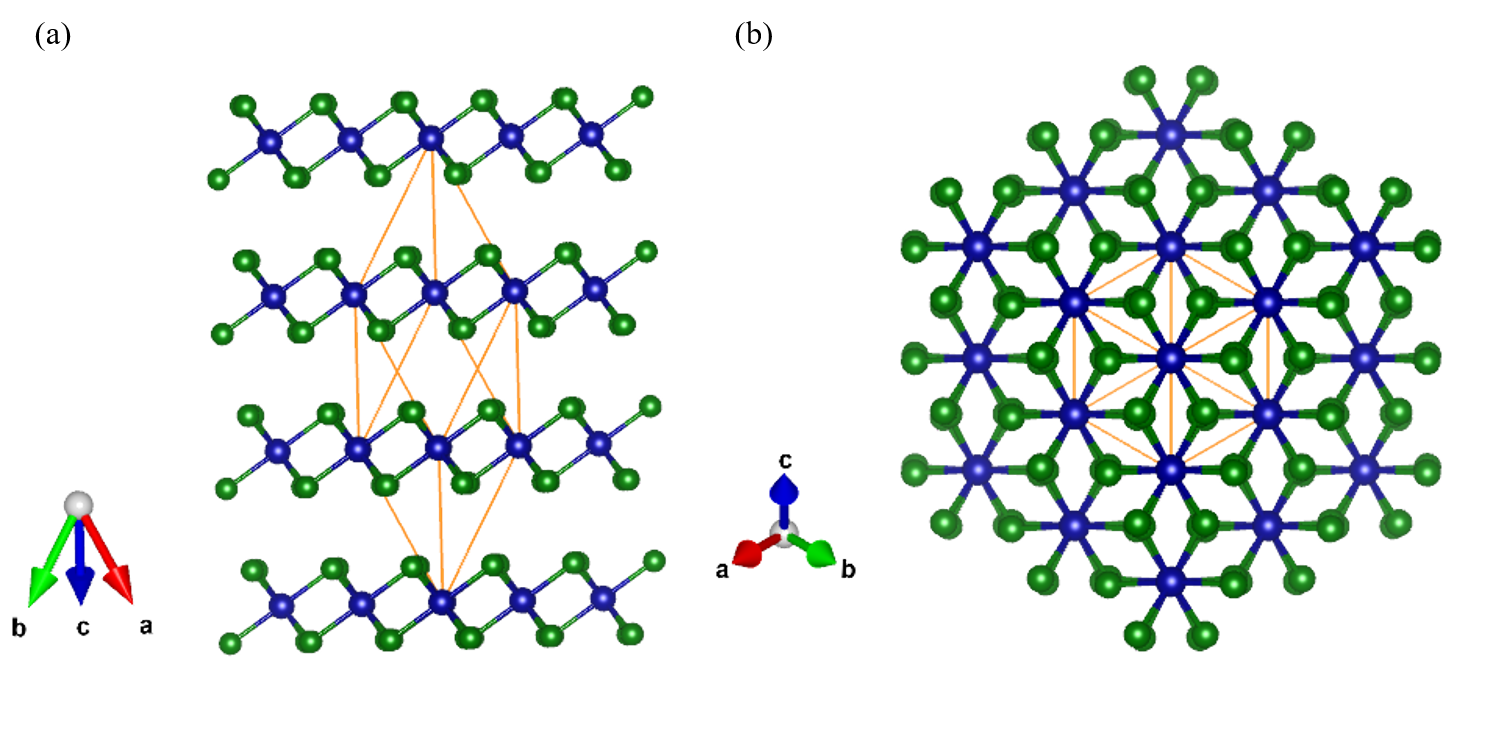}
    	\caption{The schamatic representation of (a) a side view and (b) a top view the atomic structure of the CrBr$_3$ crystal in rhombohedral configuration. The orange parallelepiped represents the unit cell, which consists of three molecular layers including two Cr atoms (blue) and six Br atoms (green) per layer.}
		\label{fig:structure}
\end{figure}

\begin{figure}[t]
		\subfloat{}%
		\centering
		\includegraphics[width=0.6\linewidth]{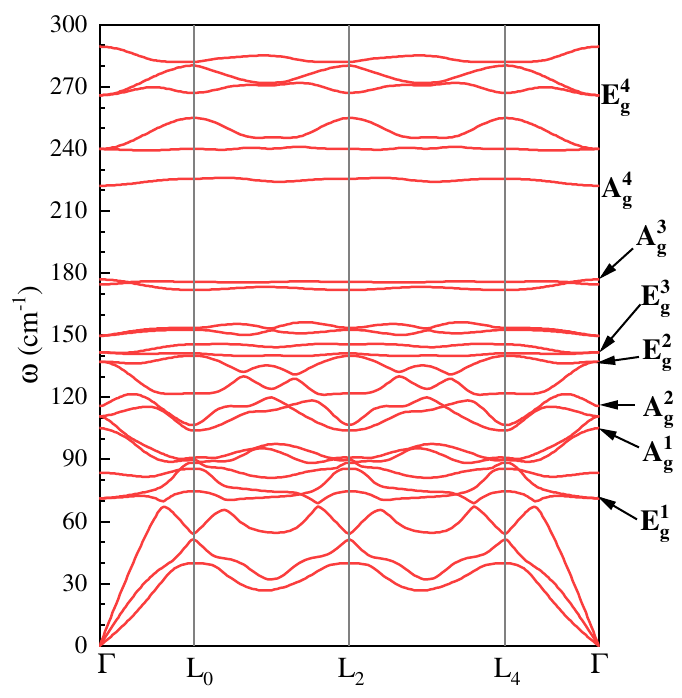}
    	\caption{Phonon dispersion of bulk CrBr$_3$ crystal in rhombohedral primitive cell with ferromagnetic order. The $\Gamma$ point corresponds to the center of the Brillouin zone.}
		\label{fig:phonon_dis}
\end{figure}

CrBr$_3$ crystallizes in a rhombohedral BI$_3$ structure of R$\bar 3$ symmetry.
The schematic representation of the atomic structure of the CrBr$_3$ crystal is shown in Fig.~\ref{fig:structure}.
Within each layer, Cr atoms form a hexagonal crystal lattice, with each atom bonded to its six neighboring Br atoms, forming an octahedral configuration.
The CrBr$_3$ layers are stacked along the $c$ axis and are held together by vdW interactions~\cite{Li2019, Craco2021, Mi2023}. 
CrBr$_3$ exibit the rhombohedral arrangement at low and ambient temperatures,~\cite{Kozlenko2021, Wu2022} unlike its counterparts, CrI$_3$ and CrCl$_3$, which undergo a structural transition from the monoclinic AlCl$_3$ phase of C2/m symmetry at ambient temperature towards rhombohedral R$\bar 3$ structure at low temperature. 
The transition was identified to occur at 210~K and 240~K for CrI$_3$ and CrCl$_3$, respectively~\cite{McGuire2015, McGuire2017}. 
The rhombohedral primitive cell belongs to space group no. 148 with Cr and Br atoms in \textit{6c} and \textit{18f} Wyckoff positions, respectively.
It gives rise to eight Raman-active modes, classified as: $\Gamma$ = 4A$_{g}$ + 4E$_{g}$, where the E$_{g}$ modes are doubly degenerate~\cite{Bermudez1976, Kozlenko2021, Wu2022}. 
We calculated the dispersion of the phonon modes for bulk CrBr$_3$ with ferromagnetic order in the density functional theory (DFT) framework, as illustrated in Fig. \ref{fig:phonon_dis}. 
The superscripts in the labels of the peaks describe additional numbering due to their increased Raman shift.
The optimized lattice constant of rhombohedral cell, $a$=7.128\AA, corresponds to $a$=6.352\AA in hexagonal cell, in good agreement with the experimental value of 6.302\AA~\cite{Kozlenko2021}.

\begin{figure}[t]
		\subfloat{}%
		\centering
		\includegraphics[width=1\linewidth]{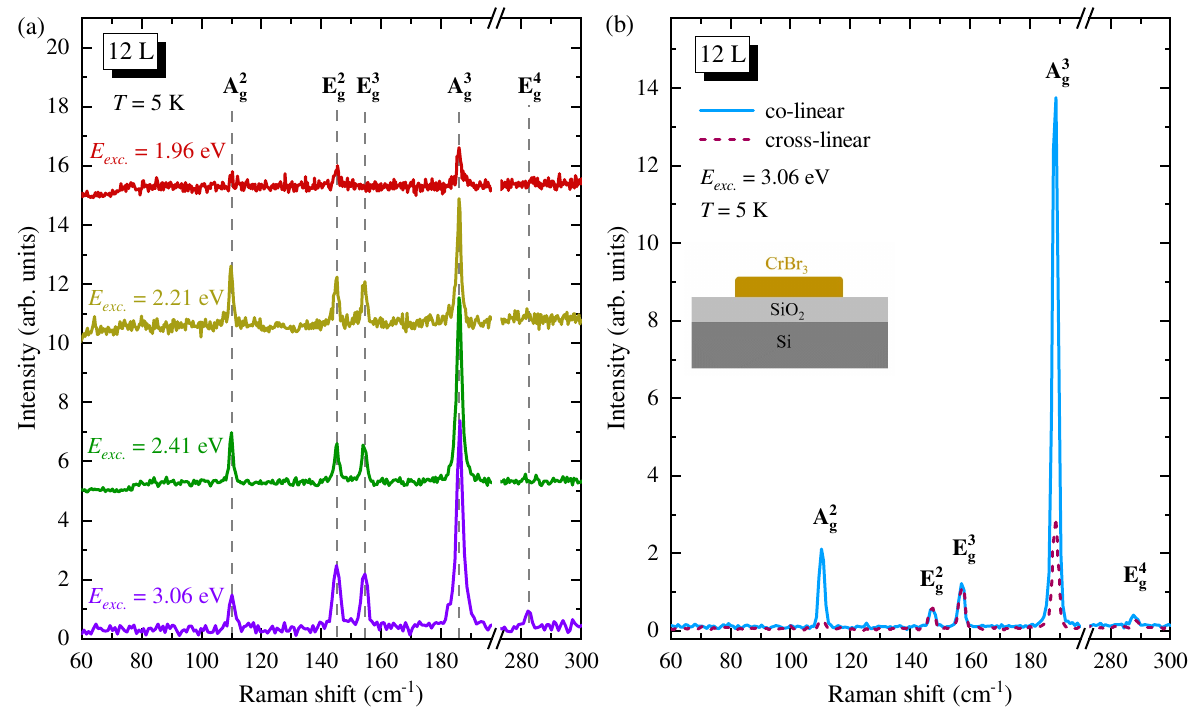}
    	\caption{(a) Raman Scattering (RS) spectra of 12 L of CrBr$_3$ measured at 5~K with different excitation energies: 1.96 eV, 2.21 eV 2.41 eV and 3.06 eV, using excitation power 50 $\mu W$. Spectra were shifted vertically for better visual representation. (b) Low-temperature ($T$~=~5~K) RS spectra of the same flake measured in co- (light-blue) and cross-linear (dashed dark pink) configurations of polarized light using 3.06 eV excitation energy and power 50 $\mu$W. Insert represents a schematic illustration of the sample cross-section.}
		\label{fig:resonant_pol}
\end{figure}

The representative low temperature ($T$=5~K) unpolarized RS spectra of a mechanically exfoliated 12~L CrBr$_3$ deposited on Si/SiO$_2$ substrate are demonstrated in Fig. \ref{fig:resonant_pol}(a). 
We comparatively inspect the dependence of the RS resonances on the laser energy, utilizing 1.96~eV, 2.21~eV, 2.41~eV, and 3.06~eV excitations. 
There are five distinct RS peaks in the spectra, and these peaks can be attributed to the in-plane E$_\textrm{g}$ and out-of-plane A$_\textrm{g}$ modes~\cite{Bermudez1976, Kozlenko2021, Yin2021, Wu2022, Lis2023}.
Specifically, four phonon modes, apparent at 110~cm$^{-1}$, 146~cm$^{-1}$, 156~cm$^{-1}$, and 187~cm$^{-1}$, are observed under all the used excitation energies, while the fifth one at 282 cm$^{-1}$ is seen exclusively when excited with a 3.06 eV energy. 
DFT calculations yield frequencies of 110~cm$^{-1}$, 137~cm$^{-1}$, 142~cm$^{-1}$, 177~cm$^{-1}$ and 266~cm$^{-1}$, respectively, which agree well with the measured values. 
We also measured polarization-resolved Raman spectra on the 12 L CrBr$_3$ flake in two configurations of linearly polarized excitation and detection: co- (light blue) and cross-linear (dashed dark pink), see Fig.~\ref{fig:resonant_pol}(b), which is consistent with the symmetry analysis of phonon modes at the $\Gamma$ point.
As can be seen in Fig. \ref{fig:resonant_pol}(a), the intensities of the Raman peaks reduce monotonically with the decrease of the excitation energy, leading to about their two times change in transition between the 1.96 eV and 3.06~eV excitations.
A similar variation in the intensity of the RS spectra under different excitations was reported in Ref.~\cite{Wu2022}, where the Raman spectra of the CrBr$_3$ monolayer were investigated using two different excitation energies, $i.e.$  1.65 eV and 2.09~eV.  
They also found that the observation of the 282 cm$^{-1}$ mode strongly depends on the used excitation energy. 

It is well established that the intensity of the RS signal in vdW materials is significantly dependent on the applied excitation energy~\cite{Zhang2015}.
In the simplest approach, the resonant conditions of RS, related to the electron-phonon coupling in a material, occur when the excitation energy of the RS processes is in the vicinity of a given transition ($e.g.$ electronic or excitonic) in the material~\cite{Carvalho2015}.
Therefore, the observed enhancement of the RS intensity measured in CrBr$_3$ under the 3.06~eV excitation should be associated with a specific transition occurring in this compound.
As the optical response of CrX$_3$ materials is determined by localized (Frenkel) excitons with large binding energies of about $\sim2-3$~eV~\cite{Acharya2021}, identification of the transition can be a difficult task.
A detailed analysis of the experimental low temperature ($T\sim2-4$~K)absorption spectra of CrBr$_3$~\cite{Kanazawa1970, Wood2004, Dillon1963, Jung1965, Dillion1966, Baral2021} along with the corresponding theoretical predictions of the imaginary part of the dielectric function~\cite{Acharya2021} reveal that CrBr$_3$ absorption is dominated by three significant resonances at energies of about 1.7~eV, 2.2~eV, and 3.0~eV. 
The highest energy absorption band apparent at about 3~eV probably is responsible for the observed resonant conditions of the RS spectra under the 3.06~eV excitation (see Fig.~\ref{fig:resonant_pol}(a)).

\begin{figure}[!t]
		\subfloat{}%
		\centering
  \includegraphics[width=1.0\linewidth]{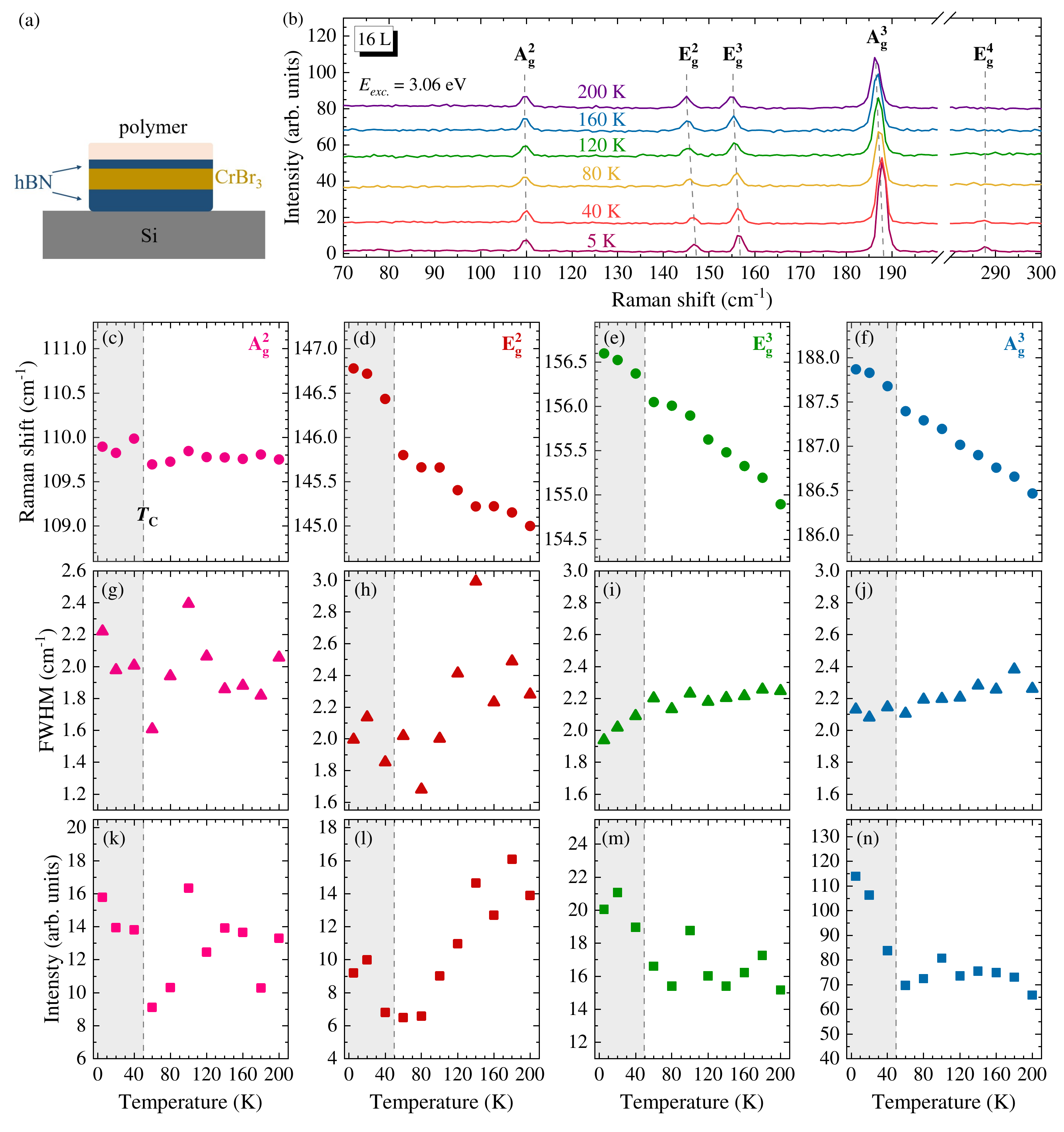}
    	\caption{(a) Schematic representation of the investigated sample. The 16 L flake of CrBr$_3$ was encapsulated in hBN and coated with an optically inactive polymer and placed on silicon substrate.
        (b) Temperature evolution of RS spectra of CrBr$_3$ measured at selected temperature between 5~K and 200~K under the 3.06~eV excitation and $P$=100~$\mu$W. 
        Temperature dependence of (c) - (f) Raman shifts, (g) - (j) FWHMs and (k) - (n) integrated intensities obtained of the A$^{1}_{\textrm{g}}$, E$^{1}_{\textrm{g}}$, E$^{2}_{\textrm{g}}$, and A$^{2}_{\textrm{g}}$ phonon modes.
        The shaded grey rectangle represents the ferromagnetic region, while the border delineates the Curie temperature ($T_{\textrm{C}}$).}
		\label{fig:temp}
\end{figure}


Most LMMs undergo different types of degradation, such as photocatalysis and photochemical or photothermal oxidation, which are substantially dependent on various factors such as light exposure and temperature~\cite{Shcherbakov2018, Wang2019, Kim2019, Tu2021, Wu2022Deg, Zhang2022}.
During our investigation, we also observed that exfoliated thin CrBr$_3$ flakes on Si/SiO$_2$ substrates are subjected to the degradation process under laser illumination, particularly at higher temperatures, close to room temperature.
Therefore, we performed the temperature-dependent RS spectra on 16 L flake encapsulated in hBN layers, and the structure was also covered by a thin layer of optically inactive polymer to add additional protection (see Methods for details) to avoid degradation (see Fig.~\ref{fig:temp}(a)).
The temperature evolution of the Raman spectra of the CrBr$_3$ flake is presented in Fig.~\ref{fig:temp}(b).
Note that we were able to carry out the measurements only up to 200~K due to the degradation caused by the elevated temperature. 
There are three effects of the increased temperature on a given Raman peak that are revealed: red shift of energy, broadening of linewidth, and variation of intensity.
To investigate the temperature effect in more detail, we fitted the observed phonon modes with Lorentzian functions. 
The obtained temperature evolutions of the peak energies, full widths at half maximum (FWHMs), and intensities for four phonon modes, $i.e.$, A$^{2}_{\textrm{g}}$, E$^{2}_{\textrm{g}}$, E$^{3}_{\textrm{g}}$, A$^{3}_{\textrm{g}}$, are shown in Figs.~\ref{fig:temp}(c)-(f), (g)-(j), and (k)-(n), respectively.

While the Raman shift of the A$^{2}_{\textrm{g}}$ mode is relatively insensitive to temperature changes, three other modes (E$^{2}_{\textrm{g}}$, E$^{3}_{\textrm{g}}$, and A$^{3}_{\textrm{g}}$) gradually redshift with increasing temperature.
Especially, in between 40~K to 60~K the energies undergo significant redshifts, followed by nearly linear decreases at higher temperatures.
These abrupt reductions were observed in the literature and identified as a fingerprint of the transition from the ferromagnetic phase at low temperature to the paramagnetic phase at higher temperatures~\cite{Bermudez1976, Kozlenko2021, Yin2021, Wu2022}.
We estimate the Curie temperature ($T_\textrm{C}$) for CrBr$_3$ to be around 50~K.
The determined $T_\textrm{C}$ value is in very good agreement with the recent result obtained for the exfoliated CrBr$_3$ flake (47~K~\cite{Yin2021}), but is much bigger value as compared to the one for a single crystal (33~K~\cite{Dillion1966} and 27~K~\cite{Kozlenko2021}) or in a monolayer limit (25~K~\cite{Wu2022}).
The difference between the found $T_\textrm{C}$ for the exfoliated flakes and the single crystal of CrBr$_3$ can be explained by the influence of stress, which can modulate the transition temperature between the ferromagnetic and paramagnetic phases~\cite{Webster2018}.

\begin{figure}[t]
		\subfloat{}%
		\centering
		\includegraphics[width=1\linewidth]{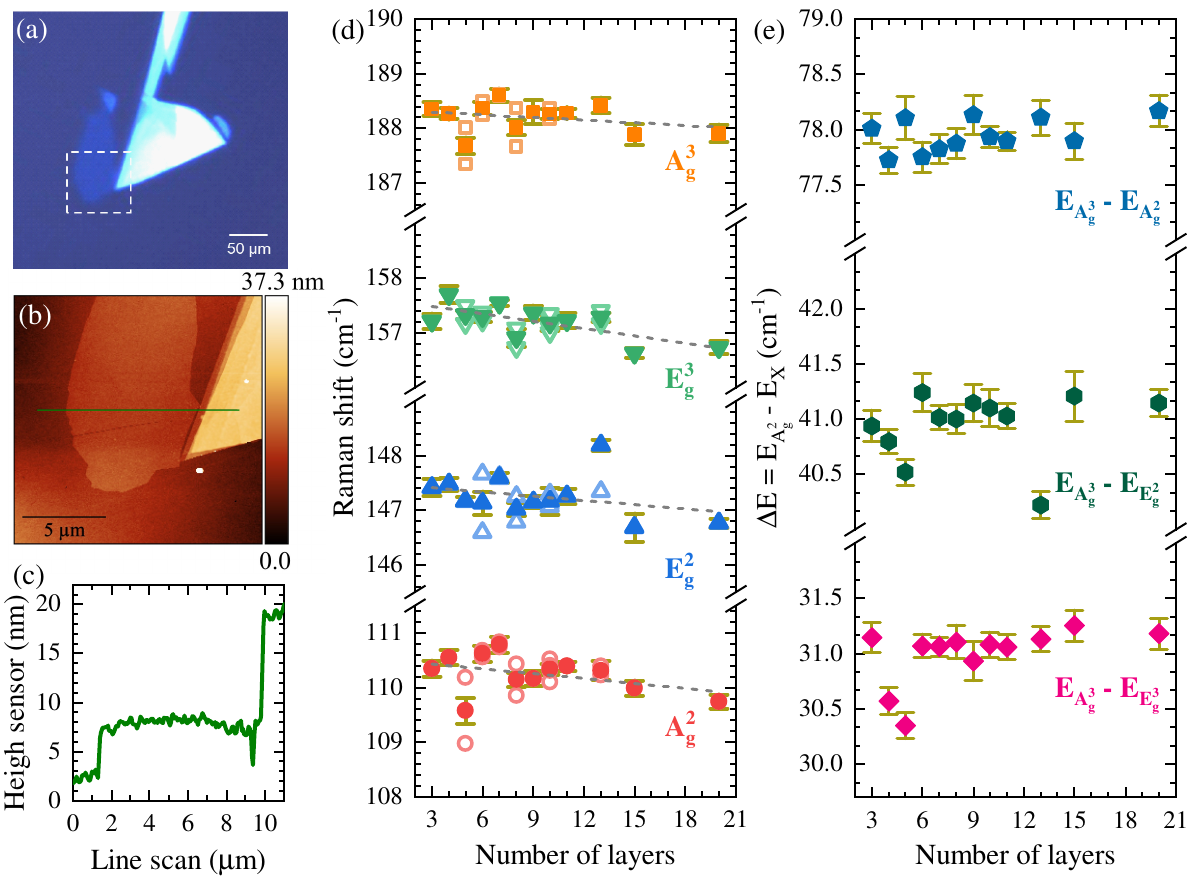}
    	\caption{(a) Optical image of 8 L flake of CrBr$_3$.
            Dashed white cube corresponds to an area scanned using AFM. 
            (b) AFM image of flake. A horizontal line represents the (c) profile of this flake. 
            (d) Raman shift as a function of number of layers. 
            The solid points represents the average Raman shift values, while the open, lighter points correspond to values of Raman shift for each layer. 
            Dark yellow bars indicate the error bars. 
            (e) Energy difference between the most prominent A$^{2}_{g}$ Raman mode and the remaining modes. 
            The dark yellow bars represents the associated error bars.}
		\label{fig:layers}
\end{figure}

For FWHMs, we observe a chaotic distribution of the FWHMs for the A$^{2}_{\textrm{g}}$ and E$^{2}_{\textrm{g}}$ modes, whereas the corresponding dependences for E$^{3}_{\textrm{g}}$, and A$^{3}_{\textrm{g}}$ are monotonic with their slow growth at higher temperatures.
There is no significant signature of the ferro-to-paramagnetism transition in the temperature evolutions of the shown FWHMs, in contrast to the results reported in Refs.~\cite{Kozlenko2021, Yin2021}.
However, Kozlenko $et$ $al.$~\cite{Kozlenko2021} demonstrated that the mode linewidths are reduced within the ferromagnetic phase, reaching a minimum at $T_\textrm{C}$ and are followed by their monotonic growth at higher temperatures.
On the other hand, Yin $et$ $al.$~\cite{Yin2021} showed similar linewidth behaviors, but the minimum of phonon FWHMs was identified at lower temperature than $T_\textrm{C}$.
Our results accompanied by the literature~\cite{Kozlenko2021, Yin2021} indicate that the analysis of the phonon linewidths in CrBr$_3$ is burdened with greater uncertainty in assigning the Curie temperature.

The analysis of the temperature dependence of the phonon intensities is the most intriguing, as there can be two competing transition processes: nonresonant versus resonant conditions of RS and ferro- versus paramagnetic order.
The intensities of the A$^{2}_{\textrm{g}}$ and E$^{3}_{\textrm{g}}$ modes exhibit a seemingly random distribution, while the corresponding ones for the E$^{2}_{\textrm{g}}$ and A$^{3}_{\textrm{g}}$ are described by more organized trends.
Particularly, the A$^{3}_{\textrm{g}}$ intensity is reduced of about 40$\%$  between 5~K and 50~K, and then its intensity stabilizes at almost a fixed level.
This result indicates that the intensity can also be used to distinguish the transition temperature between different magnetic phases, as reported for other LMMs, $e.g.$, CrSBr~\cite{Pawbake2023} and Cr$_2$Ge$_2$Te$_6$~\cite{Tian2016}.

The final part of our work is devoted to the examination of the thickness influence on the RS spectra of thin CrBr$_3$ exfoliated on the Si/SiO$_2$ substrates.
We investigate 25 flakes with thicknesses ranging from 3 L to 20 L.
Knowing that CrBr$_3$ degradation is quite slow in the ambient atmosphere, compared to its photocatalytic oxidation~\cite{Wu2022Deg}, the flakes thicknesses were determined using the AFM technique and then subjected to Raman spectroscopy characterization.
Figs.~\ref{fig:layers}(a) and (b) show the optical and atomic force microscopy (AFM) images of the selected CrBr$_3$ flake, respectively. 
Its thickness is estimated to be about 5 nm (see Fig.~\ref{fig:layers}(c)), which corresponds to 8~L (the experimental thickness of a single layer of CrBr$_3$ equals 6~${\si{\angstrom}}$~\cite{Baral2021}). 
Fig. \ref{fig:layers}(d) represents the thickness dependency of the Raman shifts of four phonon modes. 
The open points represent the phonon energies measured on the each thickness of the individual flakes, whereas the solid points denote the mean values calculated for a particular layer thickness.
A discernible tendency of the phonon energies as a function of the layer thickness is apparent in the figure.
However, the dependence of the Raman shifts, which is only from 0.4 cm$^{-1}$ to 0.8 cm$^{-1}$ in the investigated range of thicknesses, is hidden by their substantial random variations also within the same number of layers.
To better visualize the thickness effect, Fig.~\ref{fig:layers}(e) illustrates the energy differences between the most prominent A$^{3}_{g}$ peak and others (A$^{2}_{\textrm{g}}$, E$^{2}_{\textrm{g}}$, and E$^{3}_{\textrm{g}}$).
The observed dependences can be divided into two groups: 3 - 5 L and 6 -20 L.
While an evident reduction in the energy difference between the A$^{3}_{\textrm{g}}$ and E$_{\textrm{g}}$ (E$^{2}_{\textrm{g}}$ and E$^{3}_{\textrm{g}}$) modes is observed when the number of layers increases from 3 to 5 layers, there is no similar behavior for the two A$_{\textrm{g}}$ peaks (A$^{2}_{\textrm{g}}$ and A$^{3}_{\textrm{g}}$).
For the 6 - 20 L range, all the energy differences between the phonon modes are not affected by the thickness and stay almost at the same level. 
Our results, shown in Fig.~\ref{fig:layers}, are in agreement with the thickness dependences previously observed for thin layers of transition metal dichalcogenides, $e.g.$ MoS$_2$~\cite{Lee2010, Luo2013}, MoTe$_2$~\cite{Froehlicher2015, Grzeszczyk2016}, and ReSe$_2$~\cite{Kipczak2020}.
The energy differences between the A$_{\textrm{g}}$ and E$_{\textrm{g}}$ modes are established to be the result of the interaction of the interlayer and surface effects~\cite{Luo2013, Froehlicher2015}.
In the case of thin layers of CrCl$_3$, Cheng $et$ $al.$~\cite{Cheng_2021} demonstrated a significant difference in Raman shifts for 1 - 3 L and bulk. 
It may be indicative of the strong spatial localization of excitons~\cite{CrBr3_spin_pumping}, which are expected to be responsible for the resonant characteristics of the enhanced vibrational response.

\section*{Summary \label{sec:Summary}}
The vibrational and magnetic properties of the thin layers of CrBr$_3$ were investigated using Raman scattering spectroscopy.
We found that the resonant RS for CrBr$_3$ is observed at low temperature under the 3.06~eV excitation, which agrees with the reported absorption spectra of this material.
Furthermore, the temperature dependence of the phonon energies in a 20 L CrBr$_3$ encapsulated in hBN flakes was analyzed to determine its Curie temperature of about 50~K.
Finally, it was established that the effect of thickness on phonon energies is pronounced only for the thinnest layers in the vicinity of 3 to 5 layers of CrBr$_3$.

\section*{Methods \label{sec:methods}}

Most of the samples investigated in this work were thin flakes of CrBr$_3$ exfoliated directly on SiO$_2$/Si substrates with a thickness of 90 nm or 300 nm of SiO$_2$ from the bulk CrBr$_3$ crystal purchased from HQ Graphene. 
For the temperature dependence measurements, we used a sample prepared from a CrBr$_3$ flake and hBN flakes that were exfoliated directly on a 285 nm SiO$_2/Si$ substrate in an inert gas glovebox (O$_2$ < 1 ppm, H$_2$O < 1 ppm).
Then, a poly(bisphenol A carbonate)/polydimethylsiloxane, referred to as a polymer stamp on a glass slide was used to pick up 50 nm hBN, 10 nm CrBr$_3$ and 50 nm hBN, respectively at 80$^{\circ}$C with the assistance of the transfer stage in the glovebox. 
The polymer was left on the structure for protection.
CrBr$_3$ flakes of interest were first identified by visual inspection under an optical microscope and then subjected to atomic force microscopy to obtain their thickness.

Raman scattering spectra were measured under laser excitations of $\lambda$ = 405 nm (3.06~eV), $\lambda$ = 515 nm (2.41~eV), $\lambda$ = 561 nm (2.21~eV), and $\lambda$ = 633 nm (1.96~eV) on samples placed on a cold finger of a continuous-flow cryostat, which allows experiments to be carried out as a function of temperature from about 5~K to room temperature.
The excitation light was focused by means of a 50$\times$ long-working-distance objective with a 0.55 numerical aperture (NA) producing a spot of about 1 $\mu$m diameter. 
The signal was collected via the same objective (backscattering geometry), sent through a 0.75-m monochromator, and then detected using a liquid nitrogen-cooled charge-coupled device (CCD). 
The polarization-resolved Raman spectra were analyzed by a motorized half-wave plate and a fixed linear polarizer mounted in the detection path.

The first-principles calculations were performed within density functional theory (DFT) in the Vienna ab-initio simulation package (VASP)~\cite{Kresse1993, Kresse1994, Kresse1996, Kresse1996v2}. 
The projector augment wave (PAW) potentials and general gradient approximation (GGA) of Pedew-Burke-Ernzerhof (PBE)~\cite{Kresse1999} with D3 van der Waals correction~\cite{Grimme2010} were used. 
Spin-orbit coupling was not included due to its insignificant impact on lattice relaxation and phonon dispersion in this compound, as shown by Pandey $et$ $al.$ in Ref.~\cite{Pandey2022}. 
The unit cell and atomic positions have been optimized until forces on atoms were lower than 10$^{-5}$~eV/Å and stress tensor components were lower than 0.1~kbar. 
A plane wave basis set cutoff of 500~eV and a $\Gamma$-centered k-mesh 6$\times$6$\times$6 were sufficient to converge the lattice constants with precision of 0.001~\AA. 
An energy tolerance of 10$^{-8}$~eV was used to in self-consistent loop. 
The phonon dispersion was calculated using the finite displacement method as implemented in Phonopy package~\cite{Togo2023,Togo2023first}. 
2$\times$2$\times$2 supercells were found to assure convergence of the phonon frequencies at $\Gamma$ point. 
All the calculations were performed at temperature of 0~K.

\section*{Acknowledgments}
The work was supported by the National Science Centre, Poland (grant no. 2020/37/B/ST3\linebreak/02311), the Ministry of Education (Singapore) through the Research Centre of Excellence program (grant EDUN C-33-18-279-V12, I-FIM) and under its Academic Research Fund Tier 2 (MOE-T2EP50122-0012), and the Air Force Office of Scientific Research and the Office of Naval Research Global under award number FA8655-21-1-7026. The calculations were carried out with the support of the Interdisciplinary Centre for Mathematical and Computational Modelling University of Warsaw (ICM UW) under computational allocation no G95-1773.
K.W. and T.T. acknowledge support from the JSPS KAKENHI (Grant Numbers 21H05233 and 23H02052) and World Premier International Research Center Initiative (WPI), MEXT, Japan.

\section*{Author contributions}
\L{}.K., performed the measurements of the Raman scattering. R.J. and T.W. carried out DFT calculations. \L{}.K., A.K., and M.G. fabricated the samples with CrBr$_3$ thin flakes on the Si/SiO$_2$ substrates. Z.C. fabricated the sample of CrBr$_3$ encapsulated in hBN and covered by the polymer. T.T and K.W provide the hBN crystals. \L{}.K., A.K., and J.P. carried out the AFM imaging. A.B. and M.R.M. participated in the measurements of the Raman spectra. M.K. and M.R.M. supervised the project. \L{}.K. and M.R.M. analyzed the data. \L{}.K., A.K., M.G., M.K. and M.R.M wrote the manuscript with input from all coauthors.

\section*{Competing interests}
The authors declare no competing interests.

\section*{Data availability}
The datasets obtained during the experiments and analyzed for the current study are available from the corresponding authors on reasonable request.

\bibliographystyle{apsrev4-2}
\bibliography{bibio}

\end{document}